# Unveiling the charge density wave mechanism in vanadium-based Bi-layered kagome metals

Yi-Chen Yang[1,6,*], Soohyun Cho[2,*], Tong-Rui Li[2,*], Xiang-Qi Liu[3,7,*], Zheng-Tai Liu[4,*,✉], Zhi-Cheng Jiang[2], Jian-Yang Ding[1,6], Wei Xia[3,7], Zi-Cheng Tao[3,7], Jia-Yu Liu[1,6], Wen-Chuan Jing[1,6], Yu Huang[1,6], Yu-Ming Shi[1,6], Soonsang Huh[8], Takeshi Kondo[8,9], Zhe Sun[2], Ji-Shan Liu[4], Mao Ye[4], Yi-Lin Wang[5,10,11,✉], Yan-Feng Guo[3,7,✉], and Da-Wei Shen[2,✉]

[1] National Key Laboratory of Materials for Integrated Circuits, Shanghai Institute of Microsystem and Information Technology, Chinese Academy of Sciences, Shanghai, 200050, China

[2] National Synchrotron Radiation Laboratory and School of Nuclear Science and Technology, University of Science and Technology of China, Hefei, 230026, China

[3] School of Physical Science and Technology, ShanghaiTech University, Shanghai 201210, China

[4] Shanghai Synchrotron Radiation Facility, Shanghai Advanced Research Institute, Chinese Academy of Sciences, Shanghai 201210, China

[5] Hefei National Research Center for Interdisciplinary Sciences at the Microscale, University of Science and Technology of China, Hefei 230026, China

[6] University of Chinese Academy of Sciences, Beijing 100049, China

[7] ShanghaiTech Laboratory for Topological Physics, ShanghaiTech University, Shanghai 201210, China

[8] Institute for Solid State Physics, University of Tokyo, Kashiwa 277-8581, Japan

[9] Trans-scale Quantum Science Institute, University of Tokyo, Tokyo 113-0033, Japan

[10] Hefei National Laboratory, University of Science and Technology of China, Hefei 230088, China

[11] New Cornerstone Science Laboratory, University of Science and Technology of China, Hefei, 230026, China

[*] These authors contributed equally to this work

Correspondence: Da-Wei Shen (dwshen@ustc.edu.cn) or Yan-Feng Guo (guoyf@shanghaitech.edu.cn) or Yi-Lin Wang (yilinwang@ustc.edu.cn) or Zheng-Tai Liu (liuzt@sari.ac.cn)

## Abstract

The charge density wave (CDW), as a hallmark of vanadium-based kagome superconductor $AV_3Sb_5$ (A = K, Rb, Cs), has attracted intensive attention. However, the fundamental controversy regarding the underlying mechanism of CDW therein persists. Recently, the vanadium-based bi-layered kagome metal $ScV_6Sn_6$, reported to exhibit a long-range charge order below 94 K, has emerged as a promising candidate to further clarify this core issue. Here, employing micro-focusing angle-resolved photoemission spectroscopy (μ-ARPES) and first-principles calculations, we systematically studied the unique CDW order in vanadium-based bi-layered kagome metals by comparing $ScV_6Sn_6$ with its isostructural counterpart $YV_6Sn_6$, which lacks a CDW ground state. Combining ARPES data and the corresponding joint density of states (DOS), we suggest that the VHS nesting mechanism might be invalid in these materials. Besides, in $ScV_6Sn_6$, we identified multiple hybridization energy gaps resulting from CDW-induced band folding, along with an anomalous band dispersion, implying a potential electron-phonon coupling driven mechanism underlying the formation of the CDW order. Our finding not only comprehensively maps the electronic structure of V-based bi-layer kagome metals but also provide constructive experimental evidence for the unique origin of CDW in this system.

## Introduction

Kagome lattices, composed of corner-sharing triangle networks, have garnered significant research attention due to their potential for investigating the interplay between electron correlations and topologically non-trivial quantum states1[1]. Given the geometrically frustrated atomic arrangements, kagome lattices exhibit unique electronic structure characterized by Dirac-like dispersion, van Hove singularities (VHSs), and flat bands. The precise band filling of these features at the Fermi level can lead to variety of electronic instabilities and exotic quantum states, including quantum spin liquid states, superconductivity, charge or spin density waves, and Dirac or Weyl semimetals[1,2]. Very recently, a new vanadium-based kagome metal $AV_3Sb_5$ (A = K, Rb, Cs) has been discovered with remarkably similar properties to high $T_c$ superconductors, such as the competition between superconductivity and density waves, the critical scaling of the superfluid density with $T_c$ and unconventional superconducting pairing with non-trivial spin configurations[3-6]. In particular, its charge density wave (CDW) state was reported to be a unique chiral flux phase with the time-reversal symmetry breaking, which was regarded to be closely linked with unconventional superconductivity and intrinsic nematic order[5-9], and consequently kindled a wave of intense studies on the CDW in $AV_3Sb_5$. Nevertheless, despite significant progress on the origin of CDW in in $AV_3Sb_5$, the fundamental controversy between the phonon softening and electronic susceptibility instability still persists[10-14].

To elucidate the CDW mechanism in vanadium-based kagome lattices, it is essential to explore an alternative reference kagome systems. Examining similarities and differences among their CDW transitions might shed light on this issue. Recently, a new class of vanadium-based bilayer kagome metals, $RV_6Sn_6$ compounds (R = Sc, Y, ...), have been discovered analogous to the $AV_3Sb_5$. With different electron counting and negativity for rare-earth elements, $RV_6Sn_6$ exhibits intriguing and distinct properties such as the magnetic configuration, non-trivial band topology, and density waves[15-18]. Notably, $ScV_6Sn_6$ undergoes a three-dimensional CDW phase transition below $T_{CDW}$ ~ 92 K, which is accompanied by the anomalous Hall effect and breaking of the time-reversal symmetry, remarkably akin to $AV_3Sb_5$[18-22]. Moreover, this CDW can be suppressed by larger rare-earth atom substitution or pressure[23,24]. Further theoretical calculations and inelastic X-ray scattering experiments have revealed an electron-phonon coupling derived short-range CDW with $\mathbf{q_s}$ = (1/3, 1/3, 1/2), which competes with the long-range CDW order[25-27]. Besides, CDW induced pseudogap and phonon softening have been observed and discussed[26,28-31]. However, no bulk superconductivity was observed down to 80 millikelvins in $ScV_6Sn_6$, and its CDW wavevector is $\mathbf{q_{CDW}}$ = (1/3, 1/3, 1/3), both in sharp contrast to properties of $AV_3Sb_5$[7,18]. To date, the lack of a comprehensive understanding of CDWs in $ScV_6Sn_6$, particularly in the perspective of detailed low-energy electronic structure evolution upon the phase transition, significantly hinders the further exploration of charge orderings in this compound.

In this article, using micro-focusing angle-resolved photoemission spectroscopy (μ - ARPES) and first-principles calculation, we systematically investigated the underlying mechanism of CDW transition in $RV_6Sn_6$ by comparing low-energy electronic structures of Sc- and Y-compounds. Our results reveal that, although the van Hove singularities (VHSs) of $YV_6Sn_6$ is closer to the Fermi level than that of $ScV_6Sn_6$, indicative of enhanced charge instability in favor of CDW, we failed to capture any photoemission spectroscopy fingerprints of the charge order. Moreover, two-dimensional joint density of states (DOS) obtained from ARPES autocorrelation for $ScV_6Sn_6$ and $YV_6Sn_6$ both reveal an absence of CDW vector. These findings suggest that the VHSs nesting mechanism may not play a predominant role in the formation of CDW in $ScV_6Sn_6$. Interestingly, multiple instances of folded band gaps were detected within the CDW phase in $ScV_6Sn_6$, accompanied by anomalous dispersion of the band near the M-point in the Brillouin zone at a binding energy of 35 meV, indicating the potential electron-phonon coupling (EPC). Our findings not only yield direct evidence of disparities in the electronic structure within bilayer vanadium-based Kagome lattices upon substitution with distinct rare-earth ions (Sc and Y), but also provide crucial insights into the underlying origins of CDW in bilayer vanadium-based kagome metallic systems.

## Materials and Methods

### Single Crystal Synthesis

Single crystals of $RV_6Sn_6$ (R = Sc, Y) were grown via flux-based growth technique. Sc (pieces, 99%), Y (powder, 99.5%), V (powder, 99.9%), and Sn (powder, 99.99%) were loaded inside a Canfield crucible with a molar ratio of 1 : 6 : 30, sealed in a quartz tube. The sealed quartz tube loaded with the element mixtures was then heated up to 1100 °C in 15 hours in the furnace, kept at this temperature for 20 hours, and then slowly cooled down to 750 °C at a temperature decreasing rate of 2 °C/h. Single crystals were separated from the molten Sn flux via centrifuging at 750 °C. In this way, $RV_6Sn_6$ single crystals with hexagonal shiny facets in the typical size of 2 × 2 mm$^2$ were obtained.

### Angle-resolved photoemission spectroscopy

High-resolution μ - ARPES measurements were performed at 03U beamline of Shanghai Synchrotron Radiation Facility (SSRF), which is equipped with a Scienta Omicron DA30 analyzer[32]. The beam spot is around of 15 × 15 μm$^2$, and the energy and angular resolutions were set to 10 ~ 20 meV, dependent on the selected incident photon energy, and 0.2 °, respectively. The Fermi level of samples was referenced to a polycrystalline gold film mounted on the ARPES manipulator. Samples were cleaved in the ultrahigh vacuum, exposing shining (001) planes, and the pressure was maintained at less than 6 × 10$^{-11}$ Torr during measurements.

### First-principles calculation

Electronic structure of $RV_6Sn_6$ was investigated using density functional theory (DFT) within the projector augmented wave (PAW)[33] method implemented in the Vienna ab initio Simulation Package (VASP)[34]. The detailed calculation methods can be seen in Supplementary Information[35].

## Results

$ScV_6Sn_6$ crystallizes in the hexagonal centrosymmetric atomic structure with the space group P6/mmm (No. 191), as shown in Figs. 1(a) (side view) and 1(b) (top view), respectively. For our samples, the in-plane and out-of-plane lattice constants $a$ and $c$ were extracted to be 5.476 Å and 9.17 Å, respectively, consistent with previous studies[18,23]. The corresponding lattice constants of $YV_6Sn_6$ were extracted to be 5.527 Å and 9.188 Å, respectively, which do not show significant expansion, particularly for $c$, despite larger radius of Y ions. The quality of our samples has been examined by X-ray diffraction (XRD), as shown in Fig. 1(c), and the insert illustrates a typical crystal with the discernible hexagonal shape. Different from $AV_3Sb_5$, in one unit cell of $RV_6Sn_6$ there exist two discrepant kagome layers $V_3Sn$ and $SnV_3$ (originate from the slightly offset of Sn1 atoms from the V-network), which are separated by $RSn_2$ and $Sn_2$ layers alternately along the $c$ axis.

This intricate stacking generates multiple cleavage terminations, which greatly complicates ARPES measurements.

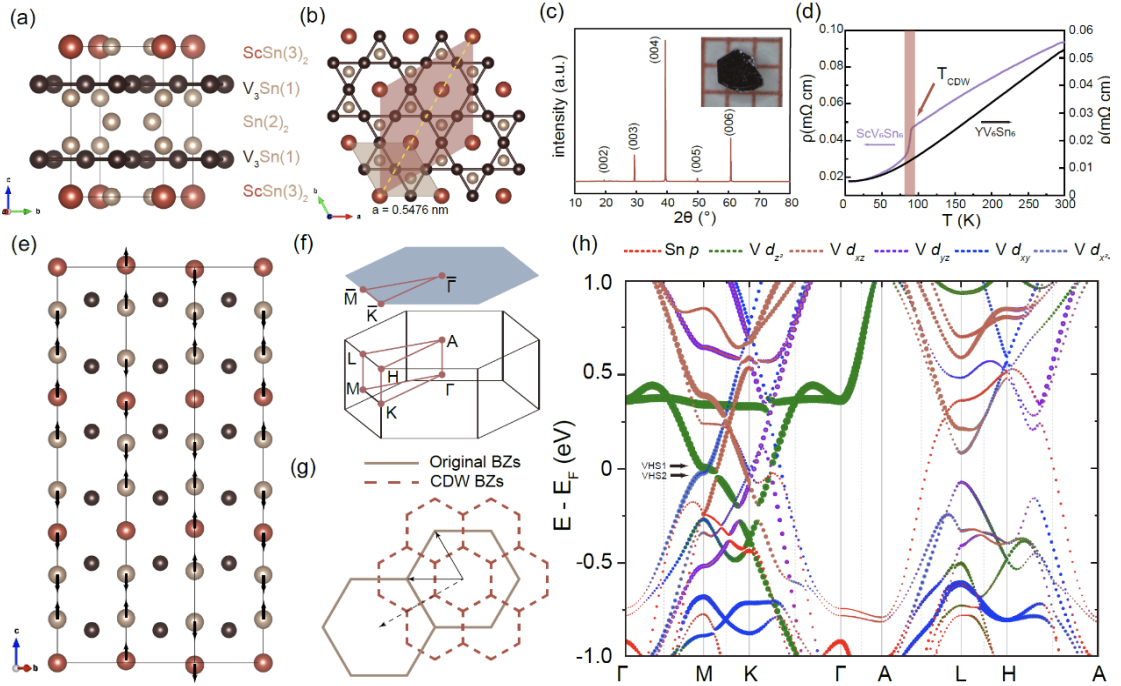

Fig. 1 (a) Side view of the layered crystal structures of $ScV_6Sn_6$. The three different kinds of Sn atoms are marked by the side number. (b) Top view of the kagome plane of $ScV_6Sn_6$. The brown and red shadow areas illustrate the pristine and ($\sqrt{3} \times \sqrt{3}$) R30° unit cell respectively. (c) The room temperature powder x-ray diffraction peaks along the *c* direction of $ScV_6Sn_6$ crystal. Inset: Optical microscope image of a typical $ScV_6Sn_6$ in this work with a well-defined hexagonal shape. (d) Temperature dependence of in-plane resistivity of the $RV_6Sn_6$ samples. The arrow indicates the anomalies associated with the CDW transition in the $ScV_6Sn_6$. (e) The side view of the reconstruction unit cell of $ScV_6Sn_6$ with ($\sqrt{3} \times \sqrt{3}$) R30° × 3 charge order along the yellow line in 1(b). The arrows indicate the modulation directions of Sc and Sn atoms along the c-axis. (f) The three-dimensional and projected two-dimensional BZs of $ScV_6Sn_6$ with marked high-symmetry points. (g) The original (brown lines) and ($\sqrt{3} \times \sqrt{3}$) R30° reconstructed (red dotted line) BZs projected on the (001) surface. (h) Orbital-projection band-structure calculation of $ScV_6Sn_6$ with spin-orbit coupling for the normal state. Different orbitals are marked by different colors. The size of the markers represents the spectral weight of the orbitals.

In Fig. 1(d), we present the comparison of temperature dependence of in-plane resistivity for $ScV_6Sn_6$ and $YV_6Sn_6$. The resistivity curve of $ScV_6Sn_6$ displays an evident anomaly upon cooling through 91 K, which was assigned to a CDW phase transition in previous reports[18,19,23,25,30,31]. In sharp contrast, $YV_6Sn_6$ does not show any anomalies in the resistivity even down to 2 K, which is coincident with the absence of CDW therein[23]. We note that, the charge ordering in $ScV_6Sn_6$ is mainly related to the modulated displacements of Sc atoms and Sn1 atoms along the *c* axis [Fig.

1(e)][18], while the V-kagome network remains nearly intact during the CDW transition, distinguishing from the inverse star-of-David CDW phase observed in AV$_3$Sb$_5$[3]. Such a lattice distortion causes a ($\sqrt{3} \times \sqrt{3}$) R30° × 3 reconstruction [Fig. 1(b)]. Accordingly, we illustrate the normal-state (original) three-dimensional and projected two-dimensional BZs and the evolution of BZs upon the CDW phase transition in Figs. 1(f) and 1(g), respectively.

To check the low-lying band structure, we firstly resorted to DFT calculations. Figure 1(h) shows the orbital-projected band-structure calculation result with spin-orbit coupling (SOC) in the normal state of ScV$_6$Sn$_6$. We found that the Fermi surface is dominated by *3d* orbitals of V, and the *5p* orbital of Sn just contributes one electron pocket around the *A* point[3]. This finding is distinct from AV$_3$Sb$_5$ in which the $p_z$ orbital of Sb is dominating in forming Fermi pockets around *Γ* and *A* points. We note that there exist two VHSs marked by VHS1 and VHS2 located in the vicinity of the Fermi level, which are attributed to different orbitals, i.e., the out-of-plane $d_{z^2}$ orbital for VHS1 and in-plane $d_{xy}/d_{x^2-y^2}$ orbital for VHS2. The proximity of these VHSs to the Fermi level is reminiscent of AV$_3$Sb$_5$, in which the charge instability induced by nesting among VHSs close to the Fermi level was reported to be the main driving force in the formation of ordered state[3,7,11,13,36]. Hereinafter, a systematic investigation of the low-lying electronic structure of ScV$_6$Sn$_6$ upon the CDW transition will be presented to address this issue.

Considering the layered structure of ScV$_6$Sn$_6$, there should exist four possible terminations, viz. two kagome terminations V$_3$Sn and SnV$_3$, one Sn$_2$ honeycomb surface and another ScSn$_2$ triangular surface. Previous ARPES studies on other layered kagome compounds have demonstrated that the formation of peaks originating from *4d* orbitals is a reliable fingerprint to distinguish different terminations[15,16,30,31]. With the assistance of μ-ARPES, we successfully established a real-space mapping of Sn *4d* photoemission intensity on the cleaved surface of ScV$_6$Sn$_6$ along the (001) crystallographic direction, as illustrated in Fig. 2(a). We could extract two kinds of distinct peak patterns for Sn *4d* states, which were attributed to the V$_3$Sn kagome termination (triple-peak feature) and ScSn$_2$ terminated surface (double-peak feature), respectively [Fig. 2(b)]. More detailed discussion regarding SnV$_3$ and Sn$_2$ terminations is provided in SI[35].

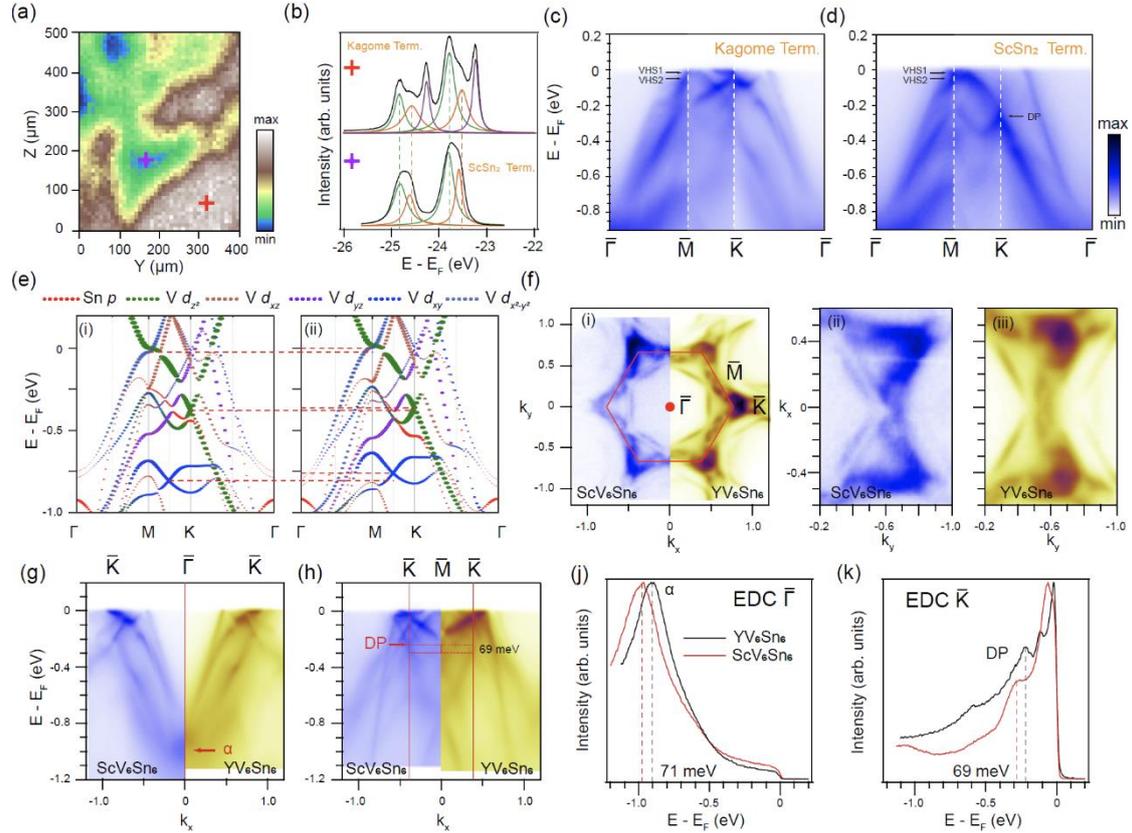

**Fig. 2 (a)** Real space mapping of Sn *4d* intensity of ScV$_6$Sn$_6$. The integration binding energy window is 23.31 ± 0.025 eV. **(b)** Sn *4d* core-level spectra for the kagome termination and ScSn$_2$ termination. **(c)(d)** Experimental ARPES *E-k* spectrum cuts along the *Γ-M-K-Γ* high symmetry direction for the V$_3$Sn Kagome termination and ScSn$_2$ termination. The typical VHS bands and Dirac cone are highlighted by the arrows. **(e)** Orbital-projection band-structure calculations of ScV$_6$Sn$_6$ and YV$_6$Sn$_6$ at k$_z$ = 0 plane. The comparison reveals that the chemical potential has a difference of 160 meV. **(f)** The comparison of Fermi surface mapping on the V$_3$Sn Kagome termination of ScV$_6$Sn$_6$ (blue) and YV$_6$Sn$_6$ (yellow). (ii) and (iii) are enlarged details around the *M* point. **(g)(h)** The intensity plots along *K-Γ-K* and *K-M-K* directions. **(j)(k)** The energy distribution curves (EDCs) at *Γ* point and *K* point, as the red lines in (g)(h)

Figures 2(c) and 2(d) show photoemission intensity plots along high-symmetry directions taken from the V$_3$Sn and ScSn$_2$ terminations at 15 K, respectively. Compared to ScSn$_2$, the photoemission result taken on V$_3$Sn kagome termination exhibits more complicated surface state around the $\bar{K}$ point (see details in SI[35]). However, we could still recognize some common bulk band features from both terminations, including the large electron-like band around $\bar{\Gamma}$, the Dirac point located at -0.21 eV at $\bar{K}$, and predicted multiple VHSs at $\bar{M}$ (VHS1 and VHS2). In general, our photoemission result shows a good agreement with the DFT calculation [Fig. 1(h)], just with a small renormalization factor of ~ 1.42, suggestive of the rather weak electron correlations therein. Figure 2(e) compares the orbital-projected band structure at k$_z$ = 0 plane with SOC between ScV$_6$Sn$_6$ (i) and YV$_6$Sn$_6$ (ii). Although Y$^{3+}$ ion processes one more electron shell than Sc, their low-energy band structure still shows remarkable coincidence except for a slight chemical potential shift. Our photoemission spectroscopy data further confirm this result. We directly compared Fermi surface maps on the V$_3$Sn kagome termination of ScV$_6$Sn$_6$ and YV$_6$Sn$_6$, as illustrated in Fig. 2(f). Their Fermi surfaces look quite similar despite the hexagonal electron pocket around $\bar{\Gamma}$ in YV$_6$Sn$_6$ shrinks slightly compared to ScV$_6$Sn$_6$, indicating an effective hole dosage therein. More comparisons of low-energy electronic structure on more terminations between these two compounds repeat such a result, as illustrated in details in Fig. S8. Figures 2(g)-2(k) display ARPES intensity plots and corresponding energy distribution curves (EDCs) at $\bar{\Gamma}$ and $\bar{K}$ on the V$_3$Sn terminations for both ScV$_6$Sn$_6$ and YV$_6$Sn$_6$. We discovered that there exists a tiny rigid band energy shift of around 70 meV for YV$_6$Sn$_6$ compared to ScV$_6$Sn$_6$.

The band energy shift observed in YV$_6$Sn$_6$ results in the convergence of VHS1 and VHS2 towards the Fermi level. The modification of VHSs is expected to enhance the charge instability and DOS of the Fermi surface, thereby favoring the formation of CDW. However, the suppression of the CDW in YV$_6$Sn$_6$ refutes the hypothesis that the VHSs nesting mechanism plays a role in the formation of charge order. The robustness of the VHSs band during the CDW phase transition provides another compelling evidence in support of our findings. In Figure S10 of the SI, we present temperature-dependent ARPES measurements of the VHSs bands, and our findings reveal no significant band renormalization. These results are consistent with some simultaneous ARPES and infrared spectroscopy studies[20,30,31].

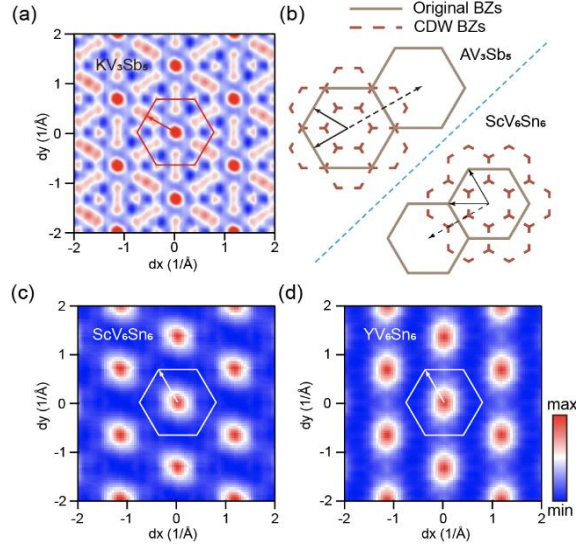

**Fig. 3 (a) Two-dimensional joint DOS results from experiment Fermi surface of KV$_3$Sb$_5$ taken at 10 K. (b) The illustration of original (brown) and CDW (dotted red) BZs of AV$_3$Sb$_5$ and ScV$_6$Sn$_6$. The continuous arrows denote the scattering vectors and the dashed arrows represent the reciprocal lattice vectors of the original BZ. (c)(d) Two-dimensional joint DOS results from experiment Fermi surface of ScV$_6$Sn$_6$ and YV$_6$Sn$_6$ taken at 10 K.**

Another potential consequence of the hole-like energy shift is the alteration in Fermi surface topology between $ScV_6Sn_6$ and $YV_6Sn_6$, which may influence the possible Fermi surface nesting. The scattering of the electrons at the Fermi surface through the phase space can be described by the two-dimensional joint DOS functions:

$$C(q) = \frac{\Omega_0}{(2\pi)^3} \int A(k, E_F)A(k+q, E_F)dk,$$

where *A(k, E_F)* is the spectral function at **E_F** at *k* point in the BZ, and $\Omega_0$ is the volume the primitive cell. One can expect the *C(q)* peaks at the ordering wave vector if there exists any charge instability due to the Fermi surface nesting. The efficacy of this joint DOS has been demonstrated in previous studies on ordered systems, such as $AV_3Sb_5$, transition metal disulfides, and cuprates[36-38]. Notably, apart from the in-site local coherence peak at **q** = 0, no additional peaks are observed throughout the BZ of both $ScV_6Sn_6$ and $YV_6Sn_6$, distinguishing them from $KV_3Sb_5$, which exhibits a peak at the CDW vector $\mathbf{q_c}$, as shown in Fig. 3. A recent calculation of the charge susceptibility function in $ScV_6Sn_6$ also highlights a distinction from $AV_3Sb_5$, as the real part of the bare charge susceptibility exhibits no peak across the entire BZ (except at the $\bar{\Gamma}$ point). In contrast, both the imaginary and real parts simultaneously exhibit peaks near the CDW vector $\mathbf{q_c}$ in $AV_3Sb_5$[27]. In conjunction with this theoretical work, our findings demonstrate that the dominance of the Fermi surface nesting mechanism in the formation of CDW in $ScV_6Sn_6$ is not supported.

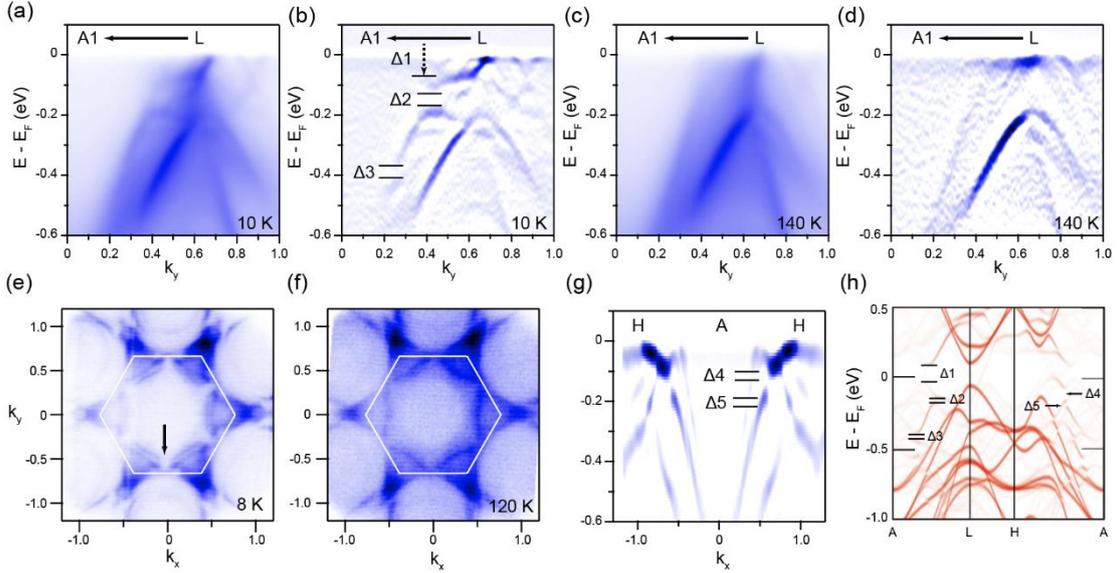

Fig. 4 (a)(b) The intensity plots and corresponding second derivative spectrum along *A - L* direction at $k_z = \pi$ at 10 K. (c)(d) Same as (a)(b) but at 140 K. (e)(f) Experimentally obtained Fermi surface map below and above the $T_{CDW}$. (g) The second derivative ARPES spectrum along the *H-A-H* direction at 8 K. (h) The unfolded band structure of the ($\sqrt{3} \times \sqrt{3}$) R30° × 3 charge order phase.

The invalidation of the VHS nesting mechanism and the rigidness of V atoms in the kagome lattices, whose orbitals dominate the low-lying electronic structure of ScV$_6$Sn$_6$, pose challenges in identifying band reconstruction in the vicinity of Fermi level therein. Nevertheless, our high-resolution ARPES measurements were still able to resolve some subtle spectral changes caused by the CDW band folding at the Sn $p_z$ orbital dominated electron band near the $A$ point in the $k_z = \pi$ plane. Figures 4(a) and 4(b) illustrate the photoemission intensity plots and corresponding second derivative spectra taken along $A$-$L$ high-symmetry direction at 10 K on the V$_3$Sn termination. Despite the inherent fragility of CDW folded bands, three band gaps (assigned as Δ1, Δ2, and Δ3, respectively) emerge as the long-range charge ordering forms. Note that spectra taken in the other high-symmetry plane, e.g., $k_z = 0$ plane, as well exhibit these gaps, as shown in Fig. S11. Deeply in occupied states, the gap Δ2 is approximately 70 meV, and Δ3 is around only 40 meV. While, Δ1 opens exactly at the Fermi level and gives rise to the Fermi pocket fracture located at $L$ [Fig. 4(e)]. Upon surpassing the critical temperature for CDW phase transition, these gaps become indiscernible and the electron Fermi pocket is reinstated to its original state [Figs. 4(c), 4(d) and 4(f)]. Similarly, spectra taken along $A$-$H$ as well exhibit such band reconstruction, as shown in Fig. 4(g). To investigate the origin of these energy gaps, we carried out the unfolded band calculation in the (√3 × √3) R30° × 3 CDW phase. The calculation presented in Fig. 4(h) clearly demonstrates that it is the hybridization between original and CDW folded bands that gives rise to these band gaps, reminiscent of the band hybridization induced gap in photoemission spectra of KV$_3$Sb$_5$ and BaFe$_2$As$_2$[7,39]. We note that such a band modification was not observed in the YV$_6$Sn$_6$ [Fig. S8(h)], in line with the absence of charge ordering therein.

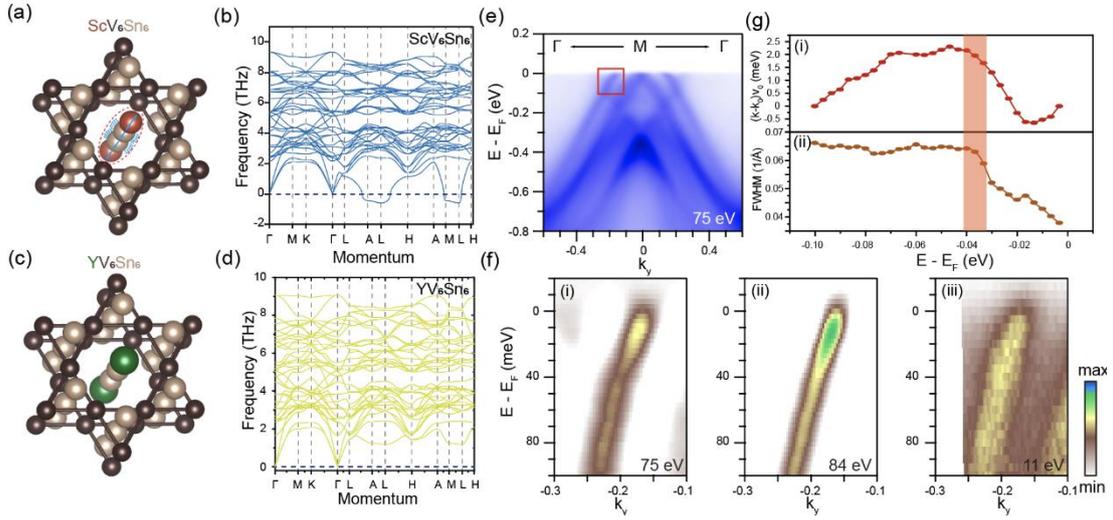

Fig. 5 (a)(b) The crystal structure of ScV$_6$Sn$_6$ that highlight the Sc-Sn-Sn-Sc atoms chain and the corresponding zero-temperature phonon spectrum. (c)(d) Same as (a)(b) but on the YV$_6$Sn$_6$. The green atoms denote the larger Y atoms. (e) The ARPES intensity on the Kagome termination of ScV$_6$Sn$_6$ along the $\Gamma$-$M$-$\Gamma$ direction. (f) The enlarged ARPES intensity plots of the band in the red box in (e) with the incident photon energies of 75 eV (i), 84 eV (ii), and 11 eV (iii). (g) (i) The peak position deviations of MDCs of the renormalized band and the bare

**band. (ii) The full-width-at-half-maximum (FWHM) of the MDC peaks. The orange shadow box shows the anomaly position.**

After eliminating the nesting induced charge instability, we next explored the phonon contribution in the CDW formation of $ScV_6Sn_6$. Figures 5(a) -5(d) show the typical one-dimensional R-Sn-Sn-R atomic chain in the Kagome frameworks and the corresponding zero-temperature phonon spectra of $ScV_6Sn_6$ and $YV_6Sn_6$. The presence of flat low-energy phonon branches in both $ScV_6Sn_6$ and $YV_6Sn_6$, as well as the observation of an imaginary frequency in $ScV_6Sn_6$, are consistent with recent theoretical studies that suggest a contribution from the longitudinal vibration of one-dimensional R-Sn-Sn-R atomic chain to these low-energy phonon modes[25-27,30,31,40]. The findings presented here offer a unique perspective for analyzing the lattice's degree of freedom: a stable Kagome framework and a one-dimensional atomic chain vibrated along the *c*-axis. These observations align with previous crystallography studies, which have demonstrated that larger R elements such as Y and Lu do not alter the out-of-plane lattice constant[23]. In the crystallographic perspective, the smaller Sc atoms can provide space to facilitate vibrations along the *c* direction, while the larger Y or Lu atom constrains these vibrations. These crystal instabilities are manifested in the low-energy branch of the phonon modes. The phonon contribution can have some traces in the spectroscopic data. The ARPES intensity plots on the Kagome termination of $ScV_6Sn_6$ in the second BZ are presented in Fig. 5(e) to mitigate the matrix element effect. In the vicinity of the Fermi level, a dispersion anomaly is observed at a binding energy of 35 meV within the hole-like band, reminiscent of kink behavior observed in $AV_3Sb_5$[13,14]. Figure 5(f) displays the enlarged spectra of this band for $ScV_6Sn_6$ with the incident photon energy of 75 eV, 84 eV, and 11 eV. This band anomaly is one of the fingerprints of electrons and bosonic modes coupling and can be quantified by the fitting of ARPES momentum distribution curves (MDCs) with a Lorentizan function. However, in contrast to the band observed in cuprate superconductors and Kagome metal $AV_3Sb_5$, this anomaly does not appear on a distinct single band, and its fitting is susceptible to interference from the intersecting electron-like surface band. Nevertheless, the peak positions obtained from the MDCs demonstrate this anomaly in $ScV_6Sn_6$, as shown in Fig. 5(g), and the FWHM (related to the imaginary part of self-energy) of the MDC peaks also has a drop at the binding energy of 35 meV. This dispersion anomaly might be the signature of the EPC in $ScV_6Sn_6$. Notably, a recent investigation using infrared spectroscopy on $ScV_6Sn_6$ has revealed an absorption peak at 34 meV[20]. The similarity in energy between this absorption peak and our observed dispersion anomaly suggests a possible relationship, although further investigation is still required.

## Discussion

In summary, we delved deeply into the formation mechanism of the CDW in the bi-layered Kagome metal $ScV_6Sn_6$. In contrast with $AV_3Sb_5$, a lot of different features indicate the invalidation of the VHSs nesting mechanism, including the mismatched charge order vector, the different atom distortion mode, the robust VHSs band during the phase transition, and the disappearance of nesting function peaks. Despite the hole-like doping in the YV$_6$Sn$_6$ pushes the VHSs to align the Fermi level, it still does not change the fact that ScV$_6$Sn$_6$ is the only exception with the CDW. Moreover, we observed the emergence of multiple energy gap opening and attributed this to the derived effect of band folding. And, anomalous band dispersion indicates the potential EPC-induced CDW signature in ScV6Sn6. Our photoemission spectroscopy investigations unveiled the minor impact of the electron interaction, highlighting instead the predominant role of the freedom of degree of the lattice.

## Acknowledgements

This work was supported by National Key R&D Program of China (Grant No. 2023YFA1406304) and National Science Foundation of China (Grant Nos. U2032208, 12004405). Y.F.G. acknowledges the Shanghai Science and Technology Innovation Action Plan (Grant No. 21JC1402000) and the open projects from State Key Laboratory of Functional Materials for Informatics (Grant No. SKL2022), CAS. Y.L.W. acknowledges the National Natural Science Foundation of China (No. 12174365) and the New Cornerstone Science Foundation. Part of this research used Beamline 03U of the Shanghai Synchrotron Radiation Facility, which is supported by ME$^2$ project under contract No. 11227902 from National Natural Science Foundation of China.

## Acknowledgements

The authors declare no competing interests.